# Noble-Metal-Assisted Fast Interfacial Oxygen Migration with Topotactic Phase Transition in Perovskite Oxides

*Qian Wang[1], Youdi Gu[2], Wenxuan Zhu[1], Lei Han[1], Feng Pan[1], and Cheng Song[1],* *

[1]Key Laboratory of Advanced Materials (MOE), School of Materials Science and Engineering, Tsinghua University, Beijing 100084, China

[2]Shenyang National Laboratory for Materials Science, Institute of Metal Research, Chinese Academy of Sciences, Shenyang 110016, China

E-mail: songcheng@mail.tsinghua.edu.cn

Transition-metal perovskite oxides constitute a series of functional material systems for electronics, catalysis and energy-conversion processes, in which oxygen migration and evolution play a key role. However, the stable metal-oxygen (*M*-O) bond forms large energy barrier inhibiting ion diffusion. Therefore, seeking efficient and facile approaches to accelerate oxygen kinetics has become a significant issue. Here, the interaction (interfacial charge transfer and cooperative bonding) between noble metal (Pt, Ag) and perovskites oxide ($SrCoO_{3-\delta}$) is employed to weaken (*M*-O) bond and decrease the energy barrier of oxygen migration. Noble metal layers serving as oxygen pumps can continuously extract oxygen from oxide films to atmosphere. The temperature of topotactic phase transition from perovskite ($SrCoO_3$) to brownmillerite ($SrCoO_{2.5}$) is remarkably lowered from ~200 °C to room temperature. Furthermore, this approach can also be applied to $SrFeO_3$ for similar topotactic phase reduction by moderate thermal activation. Our finding paves a promising and general pathway to achieve fast oxygen migration in perovskite oxides, with important application prospect in low-temperature electrodes and high-activity catalysis interface.

## 1. Introduction

Multivalent perovskite oxides show rich physical and chemical properties owing to the tunable composition, lattice structure, defects, and electronic configuration.[1-4] The control of oxygen ion/vacancy migration locates at the center of oxide-based electronics,[5–8] catalysis,[9–13] and energy fields.[3,14–16] In oxide-related energy-conversion applications, *e.g.* solid oxide fuel cells (SOFCs)[17–20] and chemical looping systems,[21–23] the $O^{2-}$ diffusion rate determines the efficiency of redox processes. However, the high thermodynamic barrier of oxygen vacancy formation due to stable metal-oxygen bonding usually prohibits fast ion diffusion.[24] Thus, high operating temperature (~700 °C) in SOFCs electrodes[25] and chemical looping[26] is usually required to guarantee reaction efficiency, which causes enormous energy dissipation. On another road, the electric field via electrolyte gating is also employed to directly drive ionic motion and even phase transition in oxides, with fast response and excellent reversibility.[27–29] Despite these advantages, voltage control still has limited change on the intrinsic ionic migration properties of oxide materials.

The basic reactions in energy-conversion process, *e.g.* oxygen evolution reaction

(OER) and oxygen reduction reaction (ORR), usually occur at the interface of oxides, which in turn affect the ion diffusion in bulk materials. The oxygen activity can be tuned by internal factors, such as covalency of metal-oxygen bond,[30] surface reconstruction,[12,31] net electronic charge of $O^{2-}$,[32] element doping,[1] *et. c*. As for external factors, the formation of oxygen vacancies can also be induced by gold nanodots decorated on $La_{0.67}Sr_{0.33}MnO_3$ surface.[33] Noble metals have a moderate interaction with oxygen and thus may activate oxygen ion in perovskite oxides, while keeping itself in free state. For transition metal oxides (*e.g.* ferrites and cobaltites), the structural and electronic characteristics are critically dependent on oxygen stoichiometry, providing an excellent platform to investigate noble metals/oxides interaction and resultant effect on oxygen migration behavior.

In this work, we choose Sr*M*$O_x$ (*M* = Co, Fe) as model systems to perform the fast oxygen migration and bias-free topotactic phase transformation from perovskite Sr*M*$O_3$ (P-S*M*O) to brownmillerite Sr*M*$O_{2.5}$ (B-S*M*O) via noble metal (Pt, Ag) decoration. During this process, interfacial charge transfer and cooperative bonding interaction play a key role in activating surficial oxygen atoms. As a convenient and effective method to achieve phase transition in perovskites, this finding provides new possibility in oxide electronics and energy-conversion applications based on accelerated interfacial oxygen kinetics assisted by noble metal layers.

## 2. Results and Discussion

SrCo$O_x$ is selected as a model system for the study of oxygen migration in perovskite oxides, due to the close relationship between oxygen stoichiometry and topotactic structure ($SrCoO_3$ in perovskite phase and $SrCoO_{2.5}$ in brownmillerite phase). The large migration barrier originates from the strong Co-O bond which has a mixed composition of covalent and ionic bonding. In the paradigm of ligand field theory,[34,35] Co 3$d$ and O 2$p$ atomic orbitals form molecular orbitals by linear combination, including bonding orbitals (1$t_{1u}$, 1$a_{1g}$, 1$e_g$), non-bonding orbitals (1$t_{2g}$) and anti-bonding orbitals (2$e_g$) in a octahedral coordination field as shown in **Figure 1**a (supposed that the $t_{2g}$ orbitals are occupied in intermediate-spin configuration[36]). The strength of Co-O bond is proportional to bond order (B.O. = ($n_b - n_a$)/2), in which $n_b$ and $n_a$ denote the number of electrons in bonding and antibonding orbitals, respectively. Therefore, electron injection to antibonding 2$e_g$ orbitals can decrease the bond order and weaken the Co-O bond strength. Here, we perform this proposal in $SrCoO_3$ by contacting with noble metal ($Nm$ = Pt, Ag) layers, in which Pt and Ag represent platinum group (Pd, Pt) and IB group (Ag, Au) at different periods in the Periodic Table.

Previous theoretical analysis about the metal/oxide interface mainly focuses on nonreducible oxides such as MgO, $TiO_2$, $SrTiO_3$,[37] in which dynamic ion migration between metal and oxides is negligible. So far, the interaction between noble metals and multivalent perovskite oxides, and the energy scale of oxygen migration dynamics have not been elucidated systematically. Here, density functional theory (DFT) calculations were used to explore the interfacial electronic structure between noble metals ($Nm$) and perovskites. Heterostructure models were constructed in two interfacial configurations: *Nm-M* (**I**) and *Nm*-O (**II**) connection (*Nm* denotes noble metal, *M* denotes transition metal cation in perovskite), in which noble metal atoms are on the top of oxygen and transition metal ions respectively (Figure S1a, Supporting Information). By comparing the energy of both structures, we found the *Nm*-O (**II**) model is more stable than *Nm-M* (**I**). *Nm* atoms preferentially adhere to surface oxygen atoms by forming *Nm*-O bond. *Nm*-O bond energy ($E_b$) can be estimated by calculating the adhesion energy ($E_a$) between $SrCoO_3$ and noble metals (Figure S1b, Supporting Information). Pt-O bond ($E_b$ = 1.416 eV) is much stronger than Ag-O bond ($E_b$ = 0.711 eV), indicating that Pt can tightly connect with oxide surfaces.

Next, differential charge densities in **Figure 1**b were calculated to display the interfacial charge transfer from noble metal to $SrCoO_3$. For both heterostructures (Pt, Ag) in *Nm*-O connection, part of the electrons moves from noble metal to the $CoO_2$ layer bridged by *Nm*-O bonds, consistent with above-mentioned Co-O bond weakening by electron injection to antibonding $2e_g$ orbitals. Due to the higher work function of Pt than Ag, the net electron donation from Pt is significantly less than that from Ag to $SrCoO_3$. Despite of this, the stronger Pt-O bond enables Pt atoms to powerfully extract oxygen from oxides. Based on theoretical calculations, **Figure 1**c designs a noble-metal-assisted deoxygenation process and resultant topotactic phase transition (TPT) from perovskite to brownmillerite phase with ordered oxygen vacancy channels (OVCs) induced by electron injection.

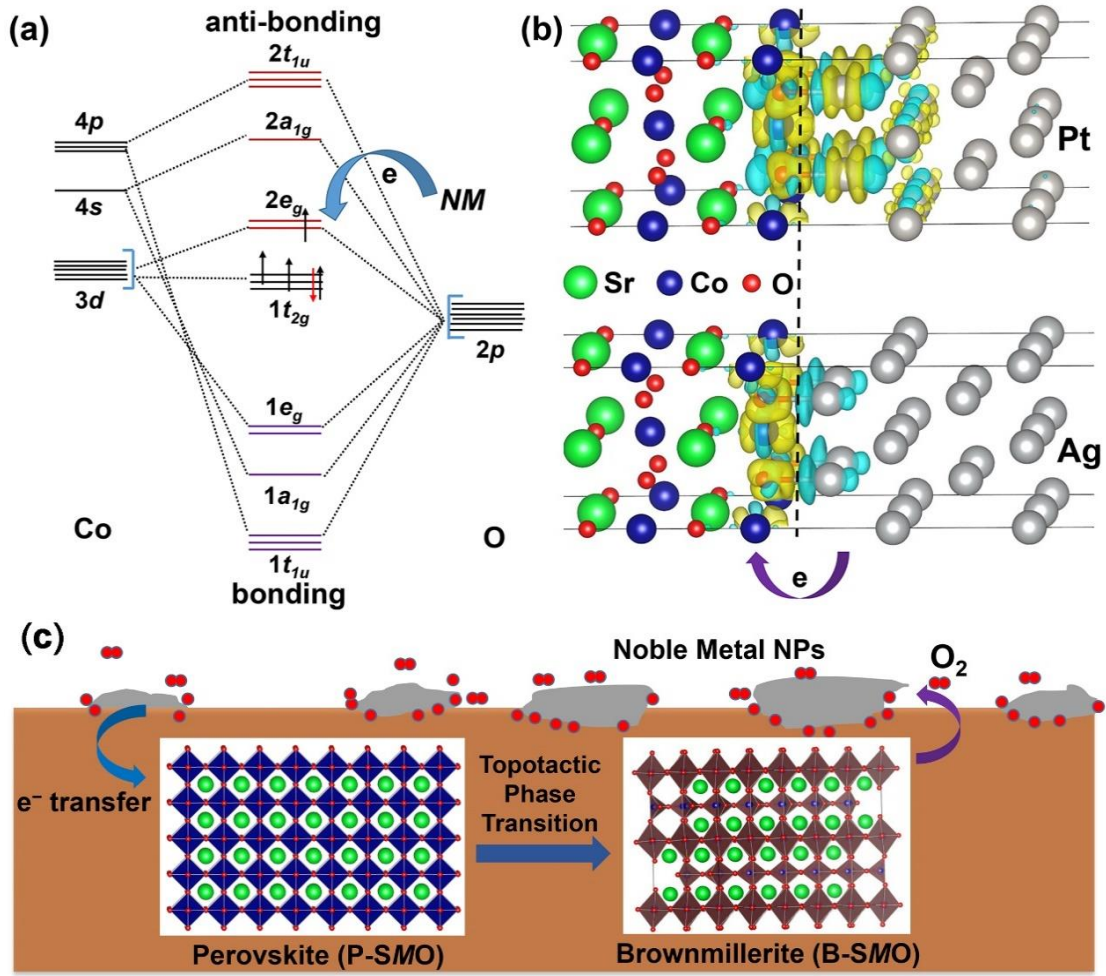

**Figure 1.** Theoretical design of barrier decrease in oxygen migration at $Nm$/ $SrCoO_3$ interface. a) molecular-orbital energy level alignment in $CoO_6$ octahedral ligand field and the principle of Co-O bond weakening by electron injection from noble metal. Bonding orbitals ($1t_{1u}$, $1a_{1g}$, $1e_g$) in low energy are fully occupied (marked in purple). Noble metal (*NM*) injects electrons to $2e_g$ antibonding orbitals to weaken Co-O bond. b) Charge transfer in noble metals/$SrCoO_3$ interface shown by differential charge density (isosurface value set to 0.003, cyan region denotes positive charge and yellow region denotes negative charge). c) Schematic diagram of deoxygenation process and topotactic phase transition (P-S*M*O to B-S*M*O) activated by noble metal nanoparticles.

Experimentally, bias-free topotactic deoxygenation process from perovskite $SrCoO_3$ (P-SCO) to brownmillerite $SrCoO_{2.5}$ (B-SCO) has been achieved at room temperature, confirming the validity of our design. Epitaxial P-SCO thin films (20~30 nm) were fabricated on $SrTiO_3$ (001) single crystal substrates by pulsed laser deposition (PLD). New-prepared P-SCO film has single perovskite phase with X-ray diffraction (XRD) peaks at 23.38° and 47.8° (**Figure 2**a), corresponding to (001) and (002) diffractions of P-SCO phase. The out-of-plane lattice constant can be determined to be $c$ = 3.80 Å (standard $c_0$ = 3.84 Å), indicating P-SCO is under the tensile strain from $SrTiO_3$ substrate. Afterwards, thin noble-metal (Pt, Ag) layers (nominal thickness ~1.8 nm) were decorated on oxide surface by electron-beam evaporation at ultrahigh vacuum. After 5 hours, the XRD spectra were measured again for $Nm$/SCO thin film to check the transition of crystal structure. **Figure 2**a gives the transformed diffraction patterns, the (001) and (002) peaks of P-SCO completely disappear for both Pt- and Ag-capped oxide films. Meanwhile, characteristic serial (0 0 $l$) peaks of oxygen-vacancy-ordered phase (B-SCO) appear at 11.36° (002), 34.24° (006), 58.78° (00<u>10</u>) respectively.[27,38,39] As a comparison, P-SCO film without $Nm$ decoration maintains invariant oxygen stoichiometry and perovskite lattice over 3 months at room temperature (Figure S2, Supporting Information), and similar TPT in ambient atmosphere requires heating (200 °C for 1 h). It should be mentioned that Pt layer grown by other method (magnetron sputtering) is also effective (Figure S3, Supporting Information), in which thermal effect in film growth can be neglected.

To explain this phenomenon from the perspective of chemical dynamics, a microscopic process of oxygen diffusion from $SrCoO_3$ to Ag was designed to explore whether the migration barrier of $O^{2-}$ can be reduced, as displayed in **Figure 2**b. The Gibbs free energy change ($\Delta G$) is −0.33 eV from first-principle calculation, which means the first step of deoxygenation process is energetically favorable due to the Ag-O affinity. In contrast, when surficial oxygen atom is moved to vacuum (*i.e.*, oxygen vacancy formation), the total energy of the system increases dramatically by +3.115 eV (Figure S4a, Supporting Information). With the existence of Ag layer, the apparent migration barrier is significantly decrease via synergistic Co-O-Ag bridged bond (cooperative bonding). Considering the incomplete spreading of Ag layer, a Ag defect site was introduced at the interface (defect-formation energy ~0.662 eV). In this case, the oxygen transfer can be easier ($\Delta G = -1.425$ eV, Figure S4b, Supporting Information), suggesting that the interfacial defect can serve as high-efficiency oxygen-transfer locations. Similar results also apply to $Pt/SrCoO_3$ model, in which the apparent dynamic barrier is decreased to +0.686 eV and −0.264 eV in perfect and Pt-defective interface, respectively (Figure S5, Supporting Information).

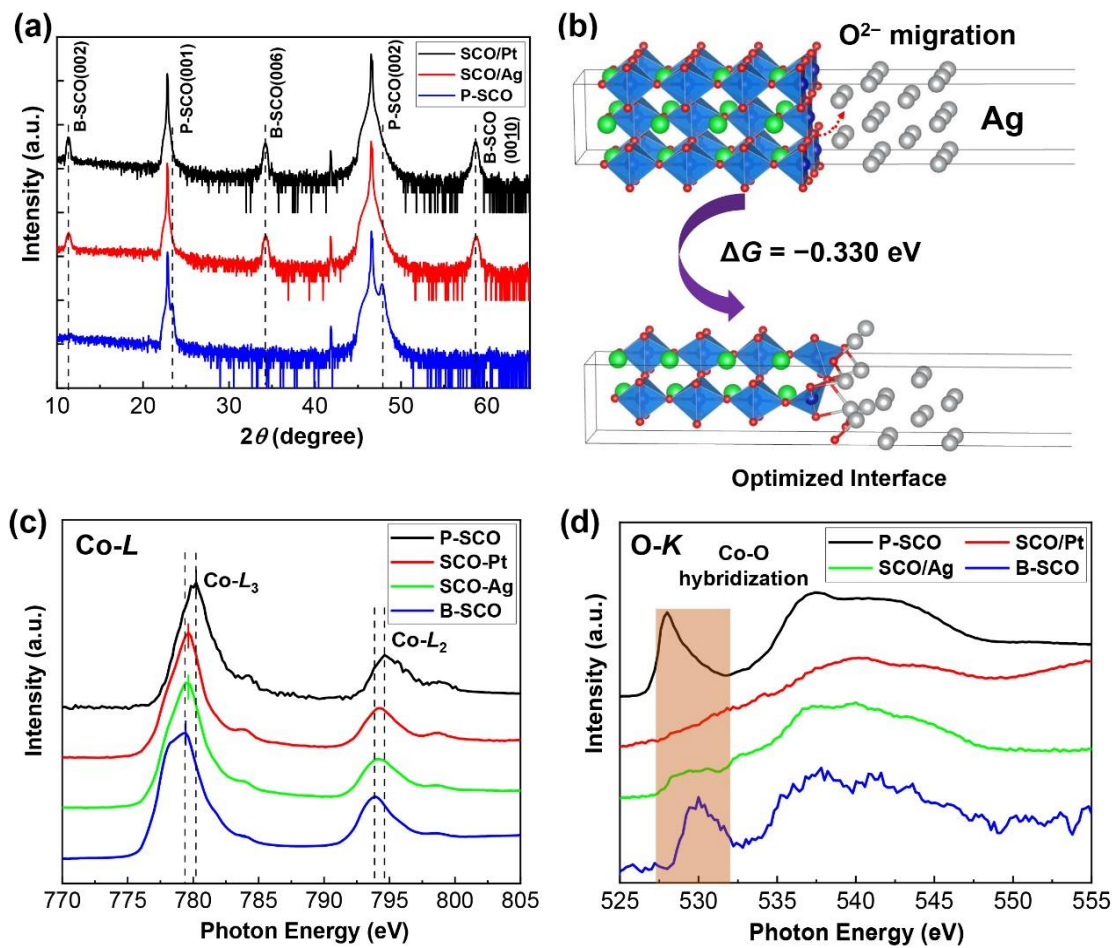

**Figure 2.** Deoxygenation-induced topotactic phase transition from P-SCO to B-SCO activated by Pt and Ag capping layers. a) X-ray diffraction $2\theta$-$\omega$ spectra of $Nm$/SCO thin film showing topotactic transition. b) Migration barrier decrease in Ag/SrCoO$_3$ heterostructure from DFT calculation. c) XAS spectra showing Co $L_{2,3}$ absorption edge shift after deoxygenation. d) The change of O $K$ absorption edge showing the weakening of Co-O hybridization. In XAS measurements, both as-grown P-SCO and B-SCO film are tested as control sample.

To substantiate that the transition originates from deoxygenation process, synchrotron X-ray absorption spectra (XAS) measurements were conducted for pristine P-SCO, Pt/SCO and Ag/SCO film, using as-grown B-SCO film as contrast. **Figure 2**c shows the Co $L_{2,3}$ absorption edge shift to lower energy after phase transition, indicating valence state reduction of Co from nominal +4 (SrCoO$_3$) to +3 (SrCoO$_{2.5}$). The coincidence of peak positions in Pt(Ag)/SCO and standard B-SCO further proves the formation of oxygen-vacancy-ordered brownmillerite phase. In addition, the remarkable change in O-$K$ edge was observed in **Figure 2**d. Both P-SCO and B-SCO have a clear pre-edge peak around 530 eV formed by Co 3$d$-O 2$p$ orbital hybridization.[27,40] Surprisingly, the O-$K$ pre-edge almost disappear in Pt(Ag)/SCO systems. Considering the thickness (1.8 nm) of noble metal layer is much smaller than detection depth (~6 nm) of XAS, this phenomenon can be attributed to the weakening of Co-O bond in the surface layer (~4 nm). The charge transfer from noble metals to SCO significantly modulates the electronic structure of surficial SCO, consistent with our theoretical calculation in **Figure 1**b. Meanwhile, the deoxygenation process induces drastic change in magnetism from ferromagnetic to antiferromagnetic property (Figure S6, Supporting Information), accompanied by the variations of electrical transport (from metallic to insulating) and optical transmittance (from dark to semitransparent) due to the suppression of Co-O hybridization and increase of bandgap (Figure S7, Supporting Information).

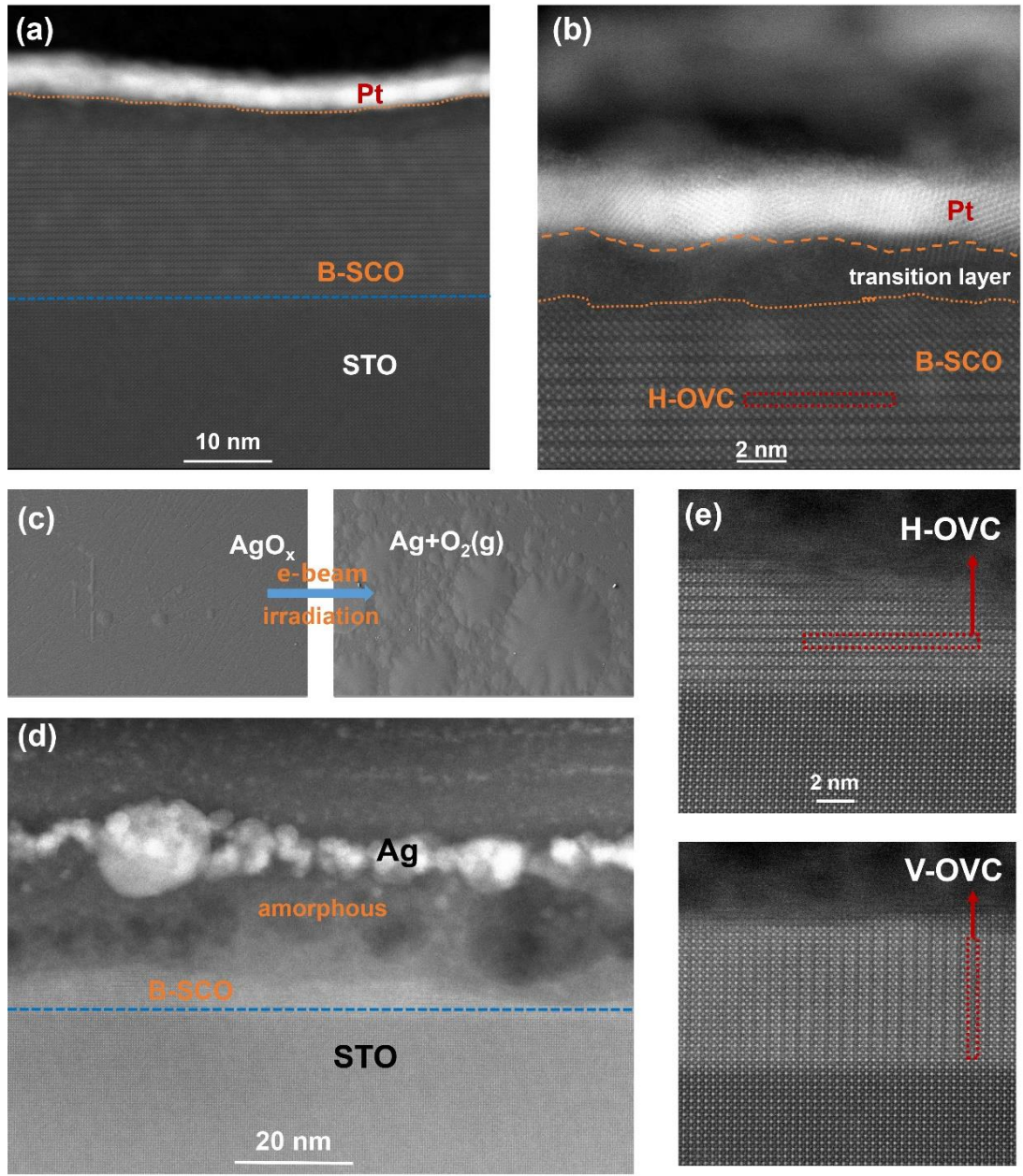

**Figure 3.** Electron microscopy analysis. a) STEM image of Pt/SCO film after phase transition. b) Enlarged structural details of Pt/SCO interface. c) Formation of oxygen bubbles below carbon film caused by Ag oxide decomposition under the irradiation of electron beam. d) STEM image of Ag/SCO film after phase transition and decomposition of $AgO_x$ layer. e) Crystalline B-SCO phase with horizontal and vertical oxygen vacancy channels (H-OVC, V-OVC) at the SCO/STO interface.

Combining structural, spectroscopic and physical-property characterizations, the bias-free oxygen migration and TPT have been solidly confirmed. To further detect the microscopic details, high angle annular dark field-scanning transmission electron microscope (HAADF-STEM) was employed to analyze Pt/SCO and Ag/SCO samples. As illustrated in **Figure 3**a, Pt layer on SCO film forms a continuous film in uniform thickness. Although atom diffusion leads to the widening of Pt layer, the interface between SCO and Pt can be clearly distinguished. P-SCO film has totally transformed to brownmillerite phase with alternate $CoO_4$ tetrahedral layers in dark contrast. The dark stripes with $2c$ period in SCO represent horizontal OVCs. Enlarged STEM image of Pt/SCO interfacial region are presented in **Figure 3**b, showing the existence of amorphous SCO transition layer between B-SCO and Pt. The destruction of crystal structure may be attributed to the fast oxygen migration at the interface which breaks up the ordered atomic arrangement of superficial SCO. The in-plane lattice constant of SCO (4.00 Å) is equal to that of STO substrate (a tiny deviation due to instrumental error), indicating SCO is perfectly epitaxial in tensile strain (Figure S8, Supporting Information).

Compared with Pt/SCO, the STEM measurement is rather difficult for Ag/SCO film. Using standard process of focused ion beam (FIB), a carbon film was first deposited on the surface of Ag/SCO for charge conduction. However, when the film was observed in scanning electron microscope (SEM) during FIB processing, numerous bubbles appeared under carbon film and gradually merged into big bulges (average size~10 μm), as displayed in **Figure 3**c. According to previous report on the reversible oxidation and reduction of Ag in TEM,[41] the gas release stems from the decomposition of Ag oxide under the irradiation of electron beam. Due to the uncontrollable release of oxygen gas, the B-SCO crystal converts to amorphous form or void (dark region in SCO film) under the disturbance of gas, as shown in **Figure 3**d. Even so, there are still a part of crystalline B-SCO (5~10 nm) remaining at SCO/STO interface. Magnified images of SCO/STO interface in **Figure 3**e display two orientations of OVCs in Ag/SCO sample, *i.e.*, horizontal (H-OVC) and vertical (V-OVC). In contrast, only H-OVC are observed in Pt/SCO sample. Although SCO films in TEM experiments are relatively thin (20~30 nm), TPT can also finish quickly for much thicker SCO films (>70 nm, Figure S3, Supporting Information). The interfacial activation effect can induce $O^{2-}$ diffusion from bulk to surface due to the chemical potential gradient of $O^{2-}$ and the formation of OVCs, which is beneficial for practical thick oxide films.

Based on STEM results, the mechanism of interfacial oxygen migration shows differences between Pt- and Ag-assisted TPT. Despite larger Pt-O bond energy, Pt keeps in metallic state due to high stability against oxidation. Therefore, Pt can only carry oxygen atoms out of SCO by exterior atoms at the grain boundary of Pt film, while the main part of Pt does not participate in oxygen migration. When P-SCO film is covered by thick Pt layer (25 nm), TPT to B-SCO can also finish within 30 min (Figure S9, Supporting Information). Therefore, thick *NM* capping do not block the emigration of oxygen, in which large-area grain boundaries and defect regions can store and transport oxygen to ambient atmosphere. As for Ag/SCO, the deoxygenation process may experience a [AgO$_x$] intermediate since the silver oxide is metastable. The decomposition of [AgO$_x$] regenerates metallic Ag, which can continuously pump oxygen from SCO matrix until the completeness of TPT. For practical applications related to oxygen migration such as SOFCs, the morphology of noble-metal layer on oxides is also dependent on external parameters and greatly influences the efficiency of charge transfer and ion diffusion.[42] Nanostructured Pt catalysts by atomic-layer deposition[43] can improve SOFC performance at low temperatures owing to a high triple-phase-boundary density[44] and electrocatalytic surface nanoionics.[45] In YSZ/Pt electrode of SOFCs, the fully mixed cermet interlayers provide thermal stability of the Pt particles and large density of catalytically active sites.[46] Our work elucidates the effect of ultrathin noble metal layers in facilitating O$^{2-}$ migration of perovskite oxides, and we speculate the activation effect can be further enhanced by maximizing the contact area of metal/oxide interface in nanoparticle or nanoporous configuration.

Furthermore, the role and evolution of noble metals in deoxygenation process were investigated by X-ray photoelectron spectroscopy (XPS). The peak position of binding energy (BE) for inner electrons reflects the oxidation states of metal elements. In this study, the Pt 4$f$ and Ag 3$d$ energy levels were taken into consideration. As control experiments, noble metals were deposited on SrTiO$_3$ and Si substrates for the same XPS measurements, in which charge transfer is relatively weak. For strongly electron-withdrawing SrCoO$_3$ film, considerable electron density was extracted away from noble metal, leading to the shift of BE to higher energy (0.09 eV for Pt 4$f$, 0.39eV for Ag 3$d$) compared with standard samples in STO and Si substrates, as shown in **Figure 4**a, b. Such differences confirm the charge-transfer indeed exists in fast oxygen-migration process. It should be mentioned here that the increase of BE also originates from saturate adsorption of oxygen atoms or molecules. When the *Nm*/SCO films were heated at 120 °C for 1 hour, the BE moves to lower energy. This result indicates the oxygen adsorbed on noble metal surface can be expelled to atmosphere by thermal effect. Thus the desorption of oxygen species from noble metal also determines the kinetics of ion migration.

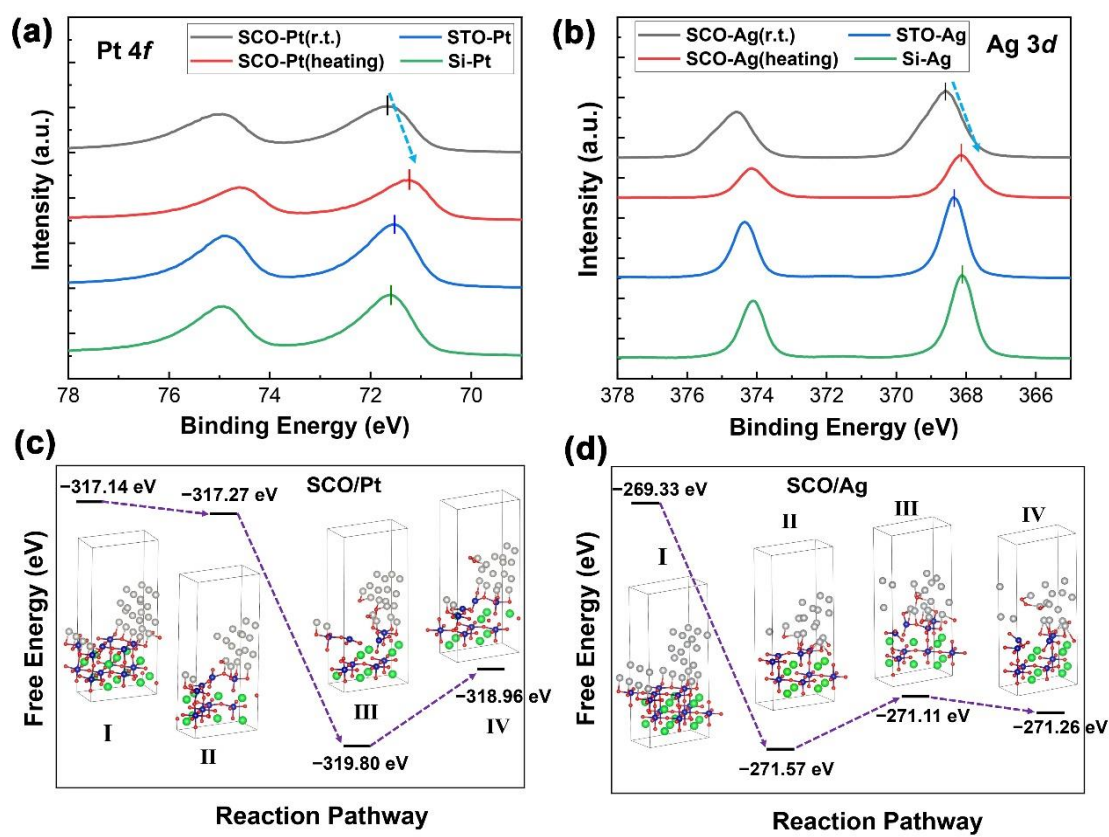

**Figure 4.** Dynamic mechanism of interfacial oxygen migration. a, b) XPS of Pt 4$f$ (a), Ag 3$d$ (b) binding energy in Pt(Ag)/SCO at room temperature, heating at 120 °C (1 h), on STO and Si substrates as contrast. c, d) Free energy evolution of Pt/SCO (c) and Ag/SCO (d) during the process of oxygen migration out of SCO film from DFT calculation.

Limited by the difficulty to characterize oxygen transport in *Nm*/SCO interface, DFT calculations were performed to explore the energy scale of oxygen migration process and illuminate the possible dynamic mechanism, as shown in **Figure 4**c, d. A simplified model was constructed using noble metal cluster (**I**) instead of heterostructure to simulate the defects in nanoparticles or grain boundary which act as active sites of oxygen evolution. Successive oxygen atom motions from SCO surface to noble metal cluster form two *Nm*-O species (**II**, **III**), accompanied by the atomic rearrangement in *Nm*/SCO interface. Next, oxygen atoms on noble metal combine into $O_2$ molecules bound in *Nm* cluster (**IV**). Initial calculation models are accessible in Figure S10 (Supporting Information). In Pt/SCO system (**Figure 4**c), the emigration of two oxygen atoms has negative free energy change (−0.13 eV and −2.53 eV) due to large Pt-O cohesive energy. The formation of Pt-O bond ($E_b$ = 1.416 eV) compensates for the energy enhancement due to the breakage of Co-O bond. However, the combination of two oxygen atoms into adsorbed $O_2$ molecules brings about obvious energy increase (+0.84 eV), originating from the reduction of Pt-O bonds. In spite of this, the energy scale is still relatively small for oxygen departure from Pt surface.

The situation in Ag/SCO system shows significant difference from that of Pt/SCO, as demonstrated in **Figure 4**d. From **I** to **II**, the free energy reduced by 2.24 eV owing to the formation of Ag-O bonds. Unexpectedly, the second oxygen atom emigration results in a slight energy enhancement by 0.46 eV, while the energy decreases again (−0.15 eV) after the two O atoms combine into $O_2$ molecule. The difference may be attributed to the small Ag-O bond energy ($E_b$ = 0.711 eV) which cannot dominate the energy variation in the whole deoxygenation process. Other factors, such as atomic rearrangement, interfacial mixing and disorder, can also induce energy fluctuation comparable with Ag-O interaction. Besides, Ag cluster is easier to change its atomic arrangement, while Pt cluster almost maintain a constant shape during the migration process. Therefore, oxygen movement in Ag layer may be more flexible than in Pt layer due to the higher mobility of Ag atom. Nevertheless, the energy evolution of both Pt/SCO and Ag/SCO shows favorable tendency for oxygen migration. On the other hand, the metastable feature of $Nm$-O species guarantees continuous oxygen pumping until the completeness of TPT, unlike active metal capping layer which cannot transport oxygen atoms once the metal is fully oxidized.[47,48]

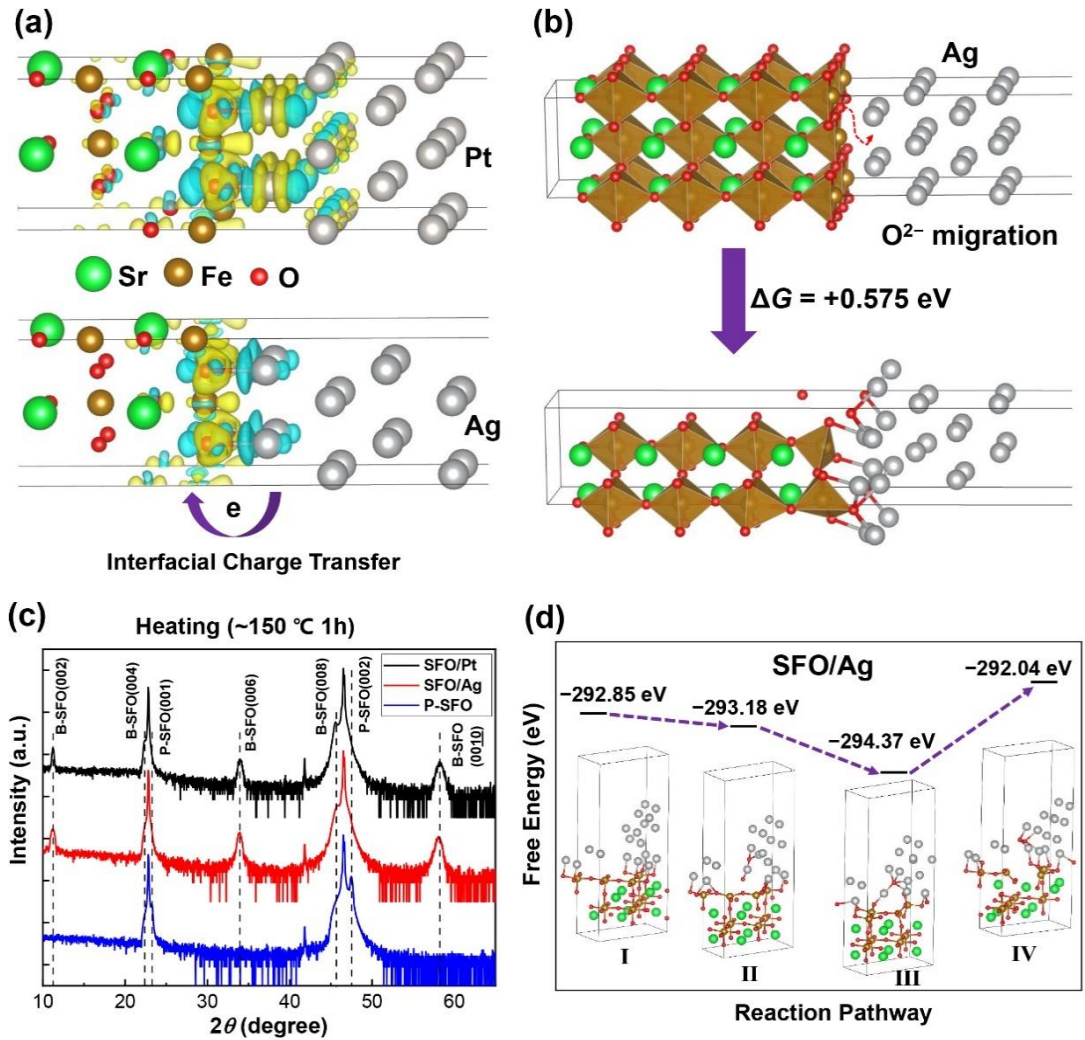

**Figure 5.** Application of noble-metal-activation method in topotactic phase transition of SrFeO$_x$. a) Differential charge density showing the interfacial charge transfer from noble metal (Pt and Ag) to SrFeO$_3$ (isosurface value set to 0.003). b) Migration barrier decrease in SFO/Ag interface. c) Noble-metal-assisted phase transition from SrFeO$_3$ (P-SFO) to SrFeO$_{2.5}$ (B-SFO) with moderate thermal activation. d) Dynamic mechanism (free energy evolution) of oxygen migration out in SFO/Ag film.

Finally, the applicability of noble-metal-assisted fast oxygen migration was extended to SrFeO$_x$ (SFO) which also has similar property of TPT,[49,50] In addition, iron-based oxides are more commonly used in SOFCs electrode and chemical looping systems,[1,22,51–53] thus the activation of oxygen ion migration in ferrites is endowed with practical significance in energy-conversion process. Similarly, we first calculated the differential charge density at the interface of $Nm$/SrFeO$_3$(P-SFO) heterostructures to see whether electron transfer happens for weakening Fe-O bond. As shown in **Figure 5**a, the situation of charge transfer from noble metal to P-SFO is analogous to $Nm$/P-SCO, indicating effective electron injection to activate stable Fe-O bond. On the other hand, the energy change of oxygen emigration to Ag atomic layer was calculated to be +0.575 eV (**Figure 5**b), much smaller than deoxygenation barrier of SrFeO$_3$ (+3.572 eV, Figure S11, Supporting Information). Likewise, Pt capping layer can also reduce the barrier of oxygen emigration to +1.339 eV (Figure S12, Supporting Information).

However, the TPT from SrFeO$_3$ to SrFeO$_{2.5}$ (B-SFO) is relatively hard compared with SCO. At room temperature, noble metal layer can only induce the formation of oxygen vacancies and the weakening of P-SFO(002) diffraction peak, but cannot form oxygen-vacancy-ordered B-SFO within several days (Figure S13, Supporting Information), due to the higher thermodynamic stability of P-SFO. Therefore, thermal effect was introduced by heating P-SFO/$Nm$ samples at ~150 °C for 1 hour. After heating, B-SFO phases appeared in both systems evidenced by the (0 0 $l$) serial diffraction peaks at 11.22 °, 22.54 °, 33.94 °, 45.58 °, 58.26 ° in **Figure 5**c, and the P-SFO (002) peak totally vanished. Heating plays two roles: first, thermodynamically reducing the free energy change to drive phase transition ($\Delta G = \Delta H - T\Delta S$, $\Delta S > 0$);

second, kinetically boosting the speed of oxygen migration at *Nm*/SFO interface and ion diffusion in the depth of films. By performing high-temperature XRD measurements for Pt/SFO sample at 150 °C, we found that the phase transition started quickly and finished within 24 min (Figure S14, Supporting Information). It is noteworthy that P-SFO keeps stable perovskite structure even at 200 °C and full phase transition needs the temperature of 300 °C (Figure S15, Supporting Information), which is greatly higher than 150 °C for noble-metal-assisted case. Furthermore, the XAS measurements on *Nm*/SFO samples (Figure S16, Supporting Information) also show the decline or disappearance of O $K$ pre-edge, indicating a weakening of Fe $3d$-O $2p$ hybridization due to electron injection and valence reduction.

The difficulty of oxygen migration in SrFeO$_3$ stems from larger deoxygenation barrier (3.572 eV) than SrCoO$_3$ (3.115 eV). Compared with Co-O bond, Fe-O bond possesses higher strength. In contrast, cobaltic (+4) compounds are less stable than similar ferric (+4) compounds. Hence, SrCoO$_3$ is easier to transform to SrCoO$_{2.5}$ with low-valence Co$^{3+}$, which is reflected by a negative Gibbs free energy change (SrCoO$_3$ → SrCoO$_{2.5}$ + 1/4 O$_2$, $\Delta G$ = −0.34 eV at 0 K) from DFT calculation. For SFO, the relative stability of P-SFO and B-SFO is determined by temperature, since the reaction (SrFeO$_3$ → SrFeO$_{2.5}$ + 1/4 O$_2$, $\Delta G$ = +0.37 eV at 0 K) is an entropy-increase process. **Figure 5**d presents an overall profile of free energy evolution in deoxygenation process of SFO/Ag system, in which the final step (oxygen atoms combination in Ag cluster) leads to a net energy enhancement of the whole process ($\Delta G$ = +0.81 eV). Therefore, TPT from P-SFO to B-SFO requires thermal activation.

Even so, the first and second step of oxygen atom emigration are energetically favorable ($\Delta G_1 = -0.33$ eV, $\Delta G_2 = -1.19$ eV), which can decrease the temperature of TPT of P-SFO to 150 °C.

## 3. Conclusion

Noble metal-metal oxide nanostructures have been extensively investigated in catalysis, solar cells, and other fields,[54] in which the interfacial coordinate bonding (*e.g.* Ti-O-Au bond in Au/TiO$_2$ nanocomposite[55]), modulation of band alignment (*e.g.* Schottky barrier and surface plasmonic resonance[56]), electronic metal-support interaction (EMSI),[57] and other interface effects play significant roles. However, the activation effects of noble metals on $O^{2-}$ migration are usually overlooked in previous studies. In this work, perovskite-oxide/noble-metal interface has been constructed to achieve fast oxygen ion migration. This effect originates from charge transfer from noble metals to oxides which weakens Co (Fe)-O bond. Additionally, noble metal layers serve as interchange of oxygen evolution via Co (Fe)−O−*Nm* cooperative bonding, which avoids the direct breakage of Co (Fe)-O bond and decrease the energy barrier. The formation of oxygen vacancies at the surface further drives the diffusion of $O^{2-}$ inside the film, and results in TPT from perovskite (P-S*M*O) to brownmillerite structure (B-S*M*O). Our work illuminates the electronic reconstruction at perovskite-oxide/noble-metal interface, paving a simple and effective way to realize fast oxygen kinetics at room temperature without external bias and high temperature. This finding will provide theoretical and practical significance for ion-migration-related catalysis and energy-conversion process based on perovskite oxides decorated by noble metals.

## 4. Experimental Section

*Film Fabrication:* Perovskite $SrCoO_{3-\delta}$ ($SrFeO_{3-\delta}$) epitaxial films were grown on (001)-oriented $SrTiO_3$(STO) substrates by pulsed laser deposition (PLD) system with 248 nm wavelength excimer laser from corresponding ceramic target. The growth parameter for $SrCoO_{3-\delta}$ ($SrFeO_{3-\delta}$) was under 720 °C (750 °C) and 210 mTorr (180 mTorr) $O_2$ atmosphere with 3 Hz (5Hz) frequency and 300 mJ (500 mJ) energy per laser pulse. After growth, annealing process for $SrCoO_{3-\delta}$ was set as fast cooling to 600 °C (650 °C) in high oxygen pressure (600 Torr) for 30 min, and then cooling down to room temperature 10 °C/min. Electron beam evaporation was used to grow Pt (Ag) layer (1.8 nm) and the growth rate was adjusted by the power of electron beam. The nominal thickness of noble metal films was determined by thin film deposition controller based on quartz crystal oscillation.

*First-principle calculation:* DFT calculations were performed in Vienna *ab initio* Simulation Packages (VASP),[58,59] using projector augmented-wave (PAW)[60] pseudopotentials with Perdew-Burke-Ernzerhof (PBE)[61] exchange-correlation functionals. Considering the strong-correlation effect of 3*d* orbitals of Co and Fe atoms,[62] GGA+*U* method[63] was applied to obtain reasonable electronic structures with appropriate effective *U* value ($U_{\text{eff}}$ = 4.0 eV). Monkhorst-Pack grids[64] were chosen as 3×3×1 (2×4×1) for structure relaxation and 9×9×1 (5×8×2) for static self-consistent and electronic structure calculations for supercell of SrCoO$_x$/*Nm* heterostructures (clusters) model. Plane-wave cutoff energy was set to 520 eV. Crystal structure data are downloaded from the Materials Project (https://materialsproject.org). Heterostructure modeling and visualization are completed in **VESTA**.[65]

*Structural Characterization:* Transition of film crystal structure was characterized by X-ray diffraction (Rigaku SmartLab X-ray Diffractometer, Cu *K*α radiation). The atomic resolution aberration-corrected scanning transmission electron microscopy (STEM) with high-angle annular dark field (HAADF) was characterized by FEI Titan Cubed Themis G2 60-300 on samples prepared by dual focused ion beam (FIB) system (Zeiss, Auriga). Morphology of film surface was measured by atomic force microscopy (AFM, Bruker Dimension FastScan).

*Physical properties measurements:* The magnetic properties of SrCoO$_x$/Ag systems were measured by a superconducting quantum interference device (SQUID) magnetometer with an in-plane magnetic field. Electrical property, resistance-temperature (*R-T*) curves were measured by physical property measurement system (PPMS, Quantum Design Inc.) in a cooling process from room temperature. Optical transmittance was measured by an ultraviolet-visible spectrophotometer (UV2600, SHIMADZU) after etching the noble metal layer by Ar$^+$ plasma.

*Spectroscopy measurements:* X-ray absorption spectra (XAS) of Cobalt $L_{2,3}$ edge and O *K* edge measurements were conducted in total electron yield (TEY) detection mode under room temperature and vacuum pressure of 8×10$^{-8}$ Torr at Beamline BL08U1A of Shanghai Synchrotron Radiation Facility (SSRF). All spectra were obtained with radiation normally incident to the film surface. The energy resolution of XAS was set to 0.2 eV, and the Co (Fe) $L_{2,3}$ edge spectra were normalized to the maximum intensities of Co (Fe) $L_3$ peak. Binding energies of Pt 4*f* and Ag 3*d* level were detected by X-ray photoelectron spectroscopy (XPS, Thermo Fisher SCIENTIFIC) with Al *K*α source gun and energy step size was set as 0.05 eV.

## Supporting Information

Supporting Information is available from the Wiley Online Library or from the author.

## Acknowledgements

Q.W. and Y.D.G. contributed equally to this work. This work was supported by the National Key Research and Development Program of China (Grant No. 2017YFB0405704), the National Natural Science Foundation of China (grant nos. 51871130 and 51671110). The authors would like to thank BL08U1A beamline at Shanghai Synchrotron Radiation Facility (SSRF) for XAS measurements. C.S. acknowledges the support of Beijing Innovation Center for Future Chip (ICFC), Tsinghua University.

## Conflict of Interest

The authors declare no conflict of interest.